\documentclass[12pt,preprint]{aastex}

\usepackage{graphicx}
\usepackage{rotating}
\shorttitle{}
\shortauthors{Nesvorn'y et al.}

\begin{document}
\baselineskip 19.pt

\title{ExoMOD II. A Statistical Model of Transit Timing Variations \\in Kepler Multi-Planet Systems}

\author{David Nesvorn\'y$^{1}$, Daniel A. Yahalomi$^{2,3}$, David Kipping$^{4}$, Cristian Beaug\'e$^{5}$,\\
Sarah C. Millholland$^{6,7}$}

\affil{(1) Solar System Science \& Exploration Division, Southwest Research Institute, 1301 Walnut Street, 
  Suite 400,  Boulder, CO 80302, USA}

\affil{(2) MIT Kavli Institute for Astrophysics and Space Research, 70 Vassar Street, Cambridge, MA 02139, USA}

\affil{(3) Juan Carlos Torres Postdoctoral Fellow}

\affil{(4) Department of Astronomy, Columbia University, 550 W 120th Street, New York, NY 10027, USA}

\affil{(5) Instituto de Astronom\'{\i}a Te\'orica y Experimental (IATE), Observatorio Astron\'omico,
Universidad Nacional de C\'ordoba, Laprida 854, X5000BGR C\'ordoba, Argentina}

\affil{(6) Department of Physics, Massachusetts Institute of Technology, Cambridge, MA 02139, USA}

\affil{(7) Kavli Institute for Astrophysics and Space Research, Massachusetts Institute of Technology, 
Cambridge, MA 02139, USA}

\begin{abstract}
  In Paper I (Nesvorn\'y et al. 2026), we forward modeled transit observations of the Kepler telescope to characterize 
  the orbital properties of close-in planetary systems. The new population model, ExoMOD, was calibrated on Kepler's 
  DR25 data. Here we use ExoMOD to statistically predict Transit Timing Variations (TTVs) from
  gravitationally interacting planets in the close-in systems, and compare these predictions with TTVs actually 
  detected in the Kepler data. We find that planet-planet interactions are not expected to produce 
  significant {\it short-period} TTVs, $P_{\rm TTV}/P_{\rm orb}<10$, where $P_{\rm TTV}$ and $P_{\rm orb}$ 
  are the TTV and orbital periods, often enough to explain the short-period TTV signals inferred from 
  the Kepler data. Most measured short-period signals must therefore have a different origin. The statistics 
  of {\it long-period} TTVs -- likely arising from the gravitational interaction between planets -- indicates that 
  single transiting planets have TTV-inducing companions nearly as often as doubles, thus ruling out multiplicity 
  distributions with a prevalence of intrinsic singles. The fraction of planets with measured long-period TTVs increases 
  to $\simeq 15$-18\% for observed multiplicities $m \geq 3$, suggesting a change in the orbital architecture. 
  High-multiplicity planetary systems have low gap complexities and probably retained a memory of their 
  formation conditions. Ultimately, our work aims at developing a more informative feedback between observations 
  and planet formation theories.  

\end{abstract}

\section{Introduction}
In a prior study (Nesvorn\'y et al. 2026; hereafter Paper I), we developed a new population model, ExoMOD, 
for Kepler's close-in planetary systems, and calibrated it on the Kepler DR25 data. ExoMOD follows the 
forward-modeling method adopted in several previous publications (e.g., Mulders et al. 2018; Zhu et al. 2018; 
He et al. 2019, 2020, 2021). First, it defines analytic representations of physical and dynamical 
properties of Kepler's close-in systems (intrinsic multiplicity, physical radius distribution of planets, 
Angular Momentum Deficit or AMD model for orbital excitation, etc.).\footnote{Following the notation
  in Paper I, we make a distinction between the {\it system} multiplicity, $m^{\rm sys}$, a system's characteristic 
  that defines the number of planets in the system, and the {\it planet} multiplicity, $m^{\rm pla}$, defined here as the 
  property of a planet indicating the number of planets in the planet's parent system. The two distributions are 
  related via ${\rm Pr}(m^{\rm pla}) \propto m^{\rm sys} {\rm Pr}(m^{\rm sys})$. We omit superscripts ``sys'' and ``pla''
  in the rest of this paper, because we always use the planet multiplicity $m^{\rm pla}$. Any $m$ appearing hereafter 
  therefore stands for $m^{\rm pla}$.} 
Second, it generates statistical samples of planetary systems and applies 
Kepler's transit-detection algorithm (Kipping \& Sandford 2016) to produce sets of model-detected planets. Third, 
model parameters are optimized by comparing the model-detected planets with the input catalog of Kepler 
planets (i.e., DR25 with several cuts: FKG dwarf hosts, planet radii $0.5 < R_{\rm pl}/R_\oplus < 7$, orbital periods 
$3<P_{\rm orb}<300$ days). The fits were executed with {\tt MultiNest} (Feroz \& Hobson 2018, Feroz et al. 2019).
See Paper I for a detailed description. Here employ ExoMOD to statistically predict Transit Timing Variations 
(TTVs; Agol et al. 2005, Holman \& Murray 2005) in Kepler multi-planet systems and compare these predictions 
with the TTV data from Kepler (Hadden \& Lithwick 2014, 2017; Holczer et al. 2016; Ofir et al. 2018; 
Yahalomi et al. 2025). 

TTVs and Transit Duration Variations (TDVs) have been widely used to characterize individual planetary systems 
(e.g., Hadden \& Lithwick 2014, 2017; Jontof-Hutter et al. 2021), including a handful of non-transiting planets (Ballard 
et al. 2011; Nesvorn\'y et al. 2012, 2013;  Dawson et al. 2014). TTV inversion is intrinsically degenerate 
when the period of the perturbing planet is unknown, as similar TTV signals can arise from a number of different 
planetary configurations (Yahalomi \& Kipping 2026), limiting unique detections in individual systems (Lammers \& Winn 2026). 
Nonetheless, TTV statistics remain informative at the population level. The basic advantage of TTVs is that 
the TTV amplitude is relatively insensitive to the mutual inclination of orbits (Nesvorn\'y 2009). A transiting  planet 
with an inclined companion will show TTVs -- assuming that the companion is massive/close enough -- independently of whether 
the companion's orbit satisfies the geometrical condition for transit. The characterization of planetary systems detected 
by the Kepler telescope (Borucki et al. 2010) can therefore be improved by adding TTVs to the list of control metrics. 

TTVs and TDVs have been used in several published statistical analyses of Kepler planets (e.g., Zhu et al. 2018, Millholland
et al. 2021, Shahaf et al. 2021, Kipping \& Yahalomi 2023, Yahalomi \& Kipping 2026). 
Here we build on these studies by: (1) taking advantage of a recent systematic re-analysis 
of periodic TTVs (Yahalomi et al. 2025) from the Holczer catalog (Holczer et al. 2016), (2) computing accurate TTVs with an 
efficient $N$-body integrator (Deck et al. 2014), and (3) integrating TTVs into a forward modeling approach where planet 
architectures can be tested against various transit and TTV constraints. For comparison, Zhu et al. (2018) only included 
$\sim$20\% of TTVs from Holczer et al. (2016) and used an approximate TTV amplitude indicator based on orbital period ratios 
of model systems (independent of planetary masses; see Appendix A).\footnote{We do not consider TDVs in this work primarily 
because we are interested in the multiplicity distribution of planets with low mutual inclinations; TDVs are typically not 
expected in these systems. TDVs and related impact parameter variations (TbDs, Judkovsky et al. 2024) are more useful in ruling 
out perturbing planets with very large orbital inclinations (Millholland et al. 2021).} Here first we describe the 
TTV catalogs (Section 2.1; Hadden \& Lithwick 2014, 2017; Holczer et al. 2016; Ofir et al. 2018; Yahalomi et al. 2025) and our 
TTV model (Section 2.2). The results are presented in Section 3. We discuss and summarize the main findings of this 
work in Sections 4 and 5.

\section{Transit Timing Variations}

\subsection{TTV catalogs} 

Holczer et al. (2016) published a transit timing catalog for KOIs from the whole Kepler mission duration. 
{They first down-sampled their KOIs from 2599 to 2339 with at least six transits. A KOI was classified as 
exhibiting significant long-term TTVs if the following statistical tests showed significance: 
(1) a modified $\chi^2$ test, indicating O-C scatter exceeded the expected timing uncertainty; 
(2) a modified power spectrum, detecting a significant periodic signal; 
(3) an alarm score, identifying correlated consecutive timing deviations; or 
(4) a polynomial fit, revealing a significant long-term trend. 
To be classified as a significant TTV signal, test (1) had to yield a $p$-value $< 10^{-4}$, the detected 
periodic variation had to have a period $P_{\rm TTV}>100$ days, and the signal had to be judged as a genuine 
astrophysical variation rather than an observational artifact. This resulted in 187 systems. Holczer et al. 
(2016) also included 73 additional KOIs that did not pass the significance threshold 
but showed convincing evidence of real long-term TTVs by eye inspection -- for a total of 260 long-term 
signals. 

Holczer et al. (2016) also highlighted 14 KOIs with significant short-term TTVs ($3<P_{\rm TTV}<80$ d). 
Ten new systems were identified by requiring a significant periodic signal in the modified power spectrum 
with a $p$-value $< 3 \times 10^{-4}$ (see Section 5 of their paper). Four additional systems
were also included: two previously reported TTVs from Mazeh et al. (2013) and two from Holczer et al. (2015). They 
pointed out that not all detected short-period modulations are due 
to physical TTVs (Szab\'o et al. 2013, Mazeh et al. 2013).} An apparent TTV periodicity can be induced 
either by the long-cadence sampling of Kepler or crossing of rotating stellar spots. For five systems,
KOI-203.01, 217.01, 883.01, 895.01 and 1074.01, the stellar rotational frequency, or its low harmonics, 
were shown to coincide with the highest TTV periodogram peak, suggesting the measured TTV periodicity was likely 
induced by stellar-spot crossing. One system, KOI-883.01, also showed a strong stroboscopic effect with 
the long-cadence sampling of Kepler. Interestingly, all these five KOIs with anomalous TTVs are in 
single-planet systems (observed multiplicity $m_{\rm obs}=1$). These considerations imply that one must be careful 
when interpreting the short-period TTVs of single-detected planets.

Yahalomi et al. (2025) re-analyzed Holczer et al. (2016) data. They aggressively removed transits
that showed any statistical sign of an anomaly and fitted, via the Lomb-Scargle (LS) periodogram, a single 
sinusoidal model to the transit times. The minimum period was set to twice the minimum sampling between 
epochs for a given KOI (the Nyquist limit). The analysis produced, for each KOI, the TTV period ($P_{\rm TTV}$), 
TTV amplitude ($A_{\rm TTV}$), and $\Delta {\rm BIC}$, where $\Delta {\rm BIC}$ is equal to the linear ephemeris 
solution Bayesian information criterion (BIC) minus the TTV solution BIC. {This is a model-selection 
approach that asks whether the improvement provided by a periodic model justifies its additional free 
parameters, rather than testing the significance of timing variations relative to a null hypothesis. 
As a result, the selection was naturally optimized for periodic, approximately sinusoidal TTVs, whereas the 
original Holczer catalog was designed to identify a broader range of TTV behaviors. Both approaches used 
the reported timing uncertainties through the $\chi^2$ likelihood, but in different ways: Holczer's criteria 
relied on $p$-values under the null hypothesis, whereas the $\Delta {\rm BIC}$ method compared the relative 
support for competing models. 
Here we favor the use of $\Delta {\rm BIC}$ over the $p$-value, because as a relative model comparison it is 
less sensitive to the absolute calibration of the uncertainties reported in Holczer et al. (2016). Specifically, 
random, non-periodic excess scatter - such as that from misestimate uncertainties - inflates $\chi^2$ of 
both the linear and periodic models comparably and largely cancels in $\Delta {\rm BIC}$, whereas the 
$p$-value responds to the absolute $\chi^2$ of the linear model and is therefore driven low by any excess 
scatter, whether or not it is periodic.}

In total, there were 1441 KOIs in Yahalomi et al. (2025) 
with a strong evidence ($\Delta {\rm BIC} > 6$) in favor of the sinusoidal TTV model over the linear ephemeris 
model.\footnote{Note that four of the five systems mentioned above, KOI-203.01, 217.01, 883.01 and 1074.01, 
were included in the joint catalog. Nearly 50\% of KOIs with significant TTVs, 687 in total in Yahalomi 
et al. (2025), have $P_{\rm TTV}<100$ d -- removing four of them would be inconsequential.} 
The dataset was linked with the Kepler input catalog described in Paper I and yielded 958 KOIs with 
significant TTVs in the ExoMOD domain.\footnote{The ExoMOD domain is defined as $\log g > 4$ (in cgs),
$0.7 < R_*/R_\odot < 1.6$, $0.5 < R_{\rm pl}/R_\oplus < 7$ and $3<P_{\rm orb}<300$ days.} 
For reasons described in Section 3, we also performed an additional analysis of the Yahalomi et al. 
(2025) data, where all short-period harmonics with $P_{\rm TTV}/P_{\rm orb}<10$ were ignored. This 
produced 502 KOIs with significant long-period TTVs in the ExoMOD domain.

Ofir et al. (2018) developed a spectral approach to TTVs. They assumed that a sinusoidal TTV exists in the 
Kepler data and calculated the chi-square improvement of this model over that of the linear-ephemeris model. 
The method enabled detection of TTVs even in cases where the transits were too shallow, such that individual 
transits could not have been timed. We downloaded the TTV catalog from Ofir et al. (2018) and cross-matched 
it with our input catalog described in Paper I, applying the same cuts as before. This resulted in 218
KOIs with significant TTVs in the ExoMOD domain. In addition, Hadden \& Lithwick (2014, 2017) obtained a much 
more strictly controlled set of planetary TTVs by selecting systems where the measured TTV period (Rowe et al. 
2015) was consistent with the super-period computed from the orbital periods of detected planet pairs.
This criterion practically guarantees that the measured TTVs are real and produced by the gravitational interaction 
between the detected planets (e.g., Steffen et al. 2013). By cross-matching these systems with our input catalog
we found 60 KOIs with significant TTVs in the ExoMOD domain; here the smaller statistics is a pay off
for confidence.  

\subsection{TTV modeling}

The preferred population models obtained in Paper I are used to generate statistical samples 
of planetary TTVs. Each planetary system with at least one model-detected planet is numerically integrated with 
a symplectic Wisdom-Holman map (Wisdom \& Holman 1991), as implemented in the code known as Swift (Levison \& 
Duncan 1994). The integration is done in Jacobi coordinates. We use the symplectic corrector from Wisdom et al. (1996).
The efficient transit detection routine in Swift was developed Nesvorn\'y et al. (2013). The code 
computes the mid-transit times by interpolation. First, the transiting planet is forward propagated on the 
ideal Keplerian orbit starting from the position and velocity recorded by Swift at the beginning of $N$-body 
time step. Second, the position and velocity at the end of the time step are propagated backward (again on 
the ideal Keplerian orbits). We then calculate a weighted mean of these two Keplerian trajectories such that 
progressively more (less) weight is given to the backward (forward) trajectory as the time approaches the
end of the time step. The transit time errors resulting from this procedure are $<1$ second.
The TTVFast code, which is widely used in exoplanet studies, implements the same scheme for transit detection 
(Deck et al. 2014). The two codes were compared in detail: they are equally fast and have similar 
accuracy.\footnote{See Deck et al. (2014) for several illustrations of model TTVs computed from TTVFast and 
Holczer et al. (2016) for TTVs obtained from the Kepler data.}  

For each transit $j$, the Swift code computes the mid-transit time $t_{{\rm mid},j}$, impact parameter $b_j$,
transit duration $T_{{\rm dur},j}$. The numerical values of $b_j$ and $T_{{\rm dur},j}$ were averaged over all
observed transits and compared to the values computed analytically (Millholland et al. 2021), 
producing consistent results. 
      
Once the precise mid-transit times are determined for all detected planets in a model system over the 
Kepler baseline, we proceed by identifying quarters in which the selected star was observed 
by Kepler. The transits occurring in the blackout quarters, where no observations of the selected star  
were available, are removed. The LS periodogram is applied to the detected transits to determine the best-fit 
single sinusoidal harmonic for each planet, thus giving us the planet's TTV amplitude and TTV period.  
The results can be compared to those obtained from the Yahalomi et al. (2025) and Ofir et al. (2018) catalogs, 
but the comparison has to be done with caution because: (1) the short-period signals in the TTV catalogs
may not be produced by the gravitational interaction between planets, 
and (2) we do not inject any photometric noise in our model, which allows us to model even very small 
TTV amplitudes ($\lesssim 1$ min) that would not be detected by Kepler (see discussion in Section 4). 
A comparison to the TTV catalog from  Hadden \& Lithwick (2014, 2017) should be less susceptible to (1).  
 
\section{TTV predictions from ExoMOD}

Paper I described several base and auxiliary models for Kepler planets. The base models adopted different 
intrinsic multiplicity distributions ($m_{\rm int}$, Poisson or Zipfian) and AMD fractions ($f_{\rm AMD}$). To define 
the AMD fraction, we determined the critical AMD, AMD$_{\rm crit}$, for the system of $m_{\rm int}$ model planets 
to be stable according to Laskar \& Petit (2017) and Petit et al. (2017). In the base models, we set
\begin{equation}
{\rm AMD}_{\rm tot} = f_{\rm AMD} {\rm AMD}_{\rm crit}\ ,
\end{equation}
where $f_{\rm AMD}$ is a fixed AMD fraction, and distributed ${\rm AMD}_{\rm tot}$ among planets following 
the algorithm described in Paper I. There were four base models in total: ${\cal M}_{219}$ with the 
Poisson distribution and $f_{\rm AMD}=1$, ${\cal M}_{220}$ with the Poisson distribution and $f_{\rm AMD}=0.3$,
${\cal M}_{221}$ with the Zipfian distribution and $f_{\rm AMD}=1$, and ${\cal M}_{222}$ with the Zipfian 
distribution and $f_{\rm AMD}=0.3$ (Table 1).

In addition, the preferred auxiliary model in Paper I, ${\cal M}_{228}$, used the Poisson multiplicity distribution and let 
{\tt MultiNest} fit for $f_{\rm AMD}$, instead of fixing it like in the base models, obtaining 
$f_{\rm AMD}=0.62\pm0.12$, a value intermediate between the one explored in the base models. As a fix 
for the gap complexity problem discussed in Paper I, the ${\cal M}_{228}$ model included a transition 
to ideally correlated orbital radii of planets for multiplicities $m_{\rm int}>m^*$, where
$m^*=4.5$ was found to best fit the Kepler data (including the gap complexity metrics). See Paper I
and Section 3 for more details. 

{The base models were designed to closely match many observed characteristics of Kepler close-in 
planets, as defined in the Kepler DR25 catalog (Christiansen et al. 2025), including: 
the (1) observed multiplicity, 
(2) orbital period distribution ($3<P_{\rm orb}<300$ d, separately for observed singles and 
observed multis, and also for the joint ensemble of planets), 
(3) orbital period ratio of neighbor planets,
(4) planet radius distribution ($0.5<R_{\rm pl}/R_\oplus<7$, where $R_\oplus$ is the Earth radius),
(5) planet radius ratio distribution of neighbor planets, 
(5) transit depth, 
(6) transit duration,
(7) period-normalized transit duration ratio distribution, and
(8) planet radius partitioning and monotonicity (separately for small planets, large planets, and 
the joint ensemble; Gilbert \& Fabrycky 2020).
In addition, the auxiliary model ${\cal M}_{228}$ also successfully matched the observed gap 
complexity (separately for observed multiplicity equal to 3, greater than 3, and also for the joint
ensemble; Gilbert \& Fabrycky 2020). The observed gap complexity distribution was not matched 
in previous population models (e.g., Mulders et al. 2018; He et al. 2019, 2020).} 

{The base and auxiliary models utilized here were selected from hundreds of models tested in Paper I. 
They are statistically favored, by a large margin, over all other tested base models. The Bayes factor 
differences of our four selected base models (Table 1) are not large enough to statistically favor any 
single one of them; that is why we include four base models in the analysis presented here. The auxiliary 
model ${\cal M}_{228}$ is statistically favored, by a large margin, over all other auxiliary models 
tested in Paper I (Table 1). The gap complexity metrics were included in the likelihood term in 
${\cal M}_{228}$, but not in the base models. We are therefore unable to statistically compare the 
base and auxiliary models based on the Bayes factor differences.} 

{In summary, five preferred population models are included in the present analysis. Given that these 
models successfully match many observed characteristics of Kepler planets (see above), we employ 
them to predict (synthetic) TTVs -- expected from the gravitational interactions of planets -- and 
ask whether these predictions match observations (e.g., measured TTV 
periods and amplitudes, correlations with observed planet multiplicity). We obtain insights 
from this study into various issues related to: (i) causes of observed TTVs (gravitational interaction 
of planets, other astrophysical phenomena, instrumental effects, etc.), (ii) TTV-informed inferences 
about the planet multiplicity distribution of Kepler planets, and, in more general terms, (iii) planet 
formation theories that suggest a significant fraction of Kepler planets may reside near orbital 
resonances (where planetary TTVs are amplified; e.g, Izidoro et al. 2017). Whereas these issues were 
already thoroughly investigated in the past, here we have the advantage to know what to expect for 
planetary TTVs from ExoMOD. We can therefore contrast these reference expectations 
with the actual TTV measurements and see what it implies for issues (i), (ii) and (iii).} 

\subsection{Short-period TTV pile-up}

\begin{figure}
\epsscale{0.8}
\plotone{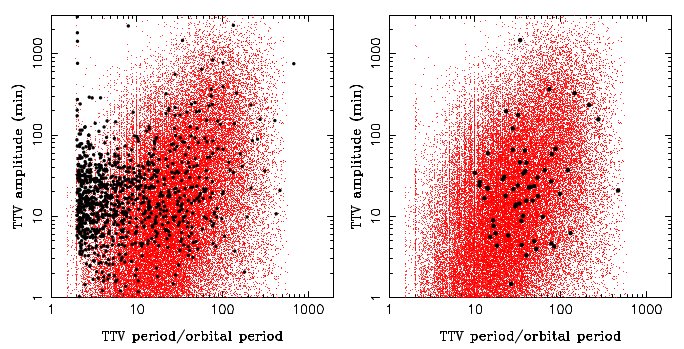}
%\epsscale{0.45}
%\plotone{fig1a.eps}\hspace*{2.mm}
%\plotone{fig1b.eps}
\caption{The TTV amplitude and TTV period normalized by the orbital period of the transiting planet, 
$P^*_{\rm TTV}=P_{\rm TTV}/P_{\rm trans}$. TTVs obtained from ${\cal M}_{219}$ are shown as red dots.
There are many model TTV points shown here because they were generated from a large ensemble of 
model systems. The black dots in the left panel are the original analysis of Holczer et al. (2016) 
TTV data from Yahalomi et al. (2025). The black dots in the right panel are the TTV data from 
Hadden \& Lithwick (2017).}
\label{scatter}
\end{figure}

We first compute TTVs from our base population models (Fig.~\ref{scatter}).
In the following text, the TTV period $P_{\rm TTV}$ is normalized by the orbital period of the transiting 
planet $P_{\rm trans}$, $P^*_{\rm TTV}=P_{\rm TTV}/P_{\rm trans}$. {The normalization of $P_{\rm TTV}$
is useful as $P_{\rm trans}$ is a natural time unit on which planetary TTVs occur, with implications
for the ``circus tent'' plot and aliasing issues discussed in Section 4. We believe that the TTV 
period normalized by the orbital period is the more relevant scaling when looking for planet-planet 
TTVs as dynamical interactions scale with the orbital period and the observed data is also dependent 
on the orbital period (one TTV data point per orbit, thus aliasing scales with the orbital period). 
We also plotted TTVs with no normalization and found that these plots do not reveal any new information 
with respect to the normalized plots; we do not show them here for brevity. The pile-up of short-period 
TTVs discussed below for $P^*_{\rm TTV}<10$ is noticeable for $P_{\rm TTV}<100$ d as well. This is not surprising  
because the median transit period in Kepler data is approximately 10 days, which means that the bulk of systems 
with $P^*_{\rm TTV}<10$ roughly corresponds to $P_{\rm TTV}<100$ d. The $P_{\rm TTV}/P_{\rm trans} = 10$ 
threshold is just a re-parameterization of the $P_{\rm TTV}>100$ day threshold (Mazeh et al. 2013, 
Holczer et al. 2016) to a more physically/dynamically motivated basis. We further address this issue
at the end of Section 3.1.}

The normalized TTV period $P^*_{\rm TTV}$ obtained in all base models shows a large spread with values up to 
$P^*_{\rm TTV} \simeq 500$ (Fig.~\ref{scatter}), 
as dictated by the shortest orbital period considered in this work, $P_{\rm orb}=3$ d, 
and the Kepler mission baseline.  
The TTV amplitudes range from $A_{\rm TTV}=1$ min, which is the lowest amplitude considered in this work,
to $A_{\rm TTV}>1000$ min. There is a hint of correlation in the population models
between $P^*_{\rm TTV}$ and $A_{\rm TTV}$ with planets having larger $P^*_{\rm TTV}$ values generally having 
larger TTV amplitudes as well. The correlation is caused by enhanced TTV amplitudes near orbital 
resonances between planets, where the super-period, and therefore the TTV period, both increase 
(Lithwick et al. 2012).\footnote{The plots with a double 
normalization, where both $P_{\rm TTV}$ and $A_{\rm TTV}$ are normalized by $P_{\rm trans}$, 
i.e., $A^*_{\rm TTV} = A_{\rm TTV}/P_{\rm trans}$, show the same general characteristics. The maximum
$A^*_{\rm TTV}$ obtained from planetary TTVs in ExoMOD follows $A^*_{\rm TTV, max} \sim 10^{-5} 
(P^*_{\rm TTV})^2$. This functional dependence is expected for low free eccentricities from Lithwick et 
al. (2012).} 
 
The original Holczer catalog analysis from Yahalomi et al. (2025) produces a pile-up of TTVs with $5<A_{\rm TTV}<100$ 
min and $2<P^*_{\rm TTV}<5$ (the left panel in Fig. \ref{scatter}). This is not expected from  
our population models where all TTVs are caused by the gravitational interaction between planets. For 
example, if we only consider the cases with $A_{\rm TTV}>10$ min, assuming that TTVs with these relatively 
large amplitudes would be detected in the Kepler data, we find from ${\cal M}_{219}$ that only $12$\% of
TTV planets should show short-periodic TTVs with $P^*_{\rm TTV}<10$ (and 88\% should be long-periodic TTVs
with $P^*_{\rm TTV}>10$). Conversely, in stark disagreement, the TTV analysis of Kepler data detects
66\% of short-periodic signals ($P^*_{\rm TTV}<10$) and only 34\% of long-periodic signals ($P^*_{\rm TTV}>10$). 
The pile-up in Kepler TTV data happens for all planets independently of the observed multiplicity of 
systems to which they belong, but there is a trend that the fraction of short-period signals decreases 
with the observed multiplicity (from 70\% for the observed multiplicity $m_{\rm obs}=1$ down to 51\% 
for $m_{\rm obs}=4$).   

We already noted in Section 2.1 that one has to be careful in interpreting the short-period TTVs,
especially for the low multiplicity systems, because not all detected short-period modulations are caused by 
dynamical TTVs. An apparent TTV periodicity can be induced, for example, by crossing of rotating stellar 
spots (Szab\'o et al. 2013, Mazeh et al. 2013). We therefore suggest that the pile-up of short-period
TTVs in Yahalomi et al. (2025) is unrelated to the gravitational interaction between planets. 

\begin{figure}
\epsscale{0.5}
\plotone{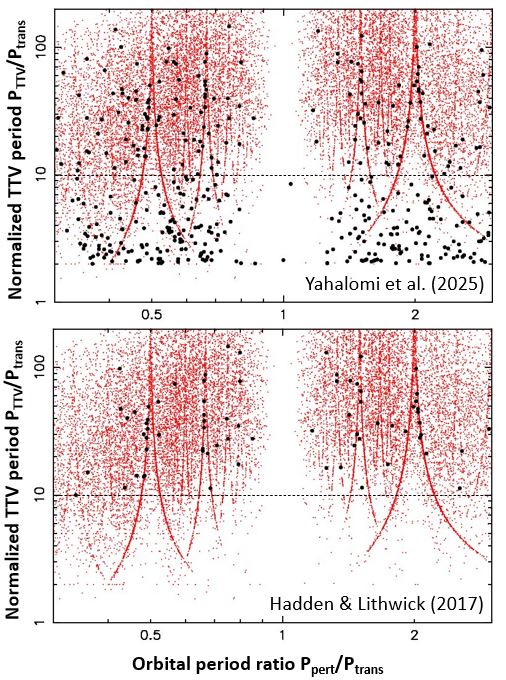}
%\epsscale{0.6}
%\plotone{circus.eps}\vspace*{5.mm}
%\plotone{circus2.eps}
\caption{Following Yahalomi et al. (2025), we plot the normalized TTV period, $P^*_{\rm orb}=P_{\rm TTV}/P_{\rm trans}$,
where $P_{\rm trans}$ is the orbital period of a transiting planet with the TTV period $P_{\rm TTV}$, 
with respect to the orbital period ratio $P_{\rm pert}/P_{\rm trans}$, where $P_{\rm pert}$ is the perturbing's planet 
orbital period (i.e., the one causing the measured TTVs). The red dots, which follow the circus 
tent profile at resonances, as expected from Yahalomi et al. (2025), are obtained from ${\cal M}_{219}$. 
This generalizes the original result of Yahalomi et al. (2025) to systems with multiplicities higher than two.  
The black dots in the top panel are from the original analysis of Holczer et al. (2016) TTV
data from Yahalomi et al. (2025). The black dots in the bottom panel show the TTV data from Hadden \& 
Lithwick (2017).}
\label{circus}
\end{figure}

We find two additional 
supporting arguments for this interpretation. First, when we plot TTVs from Hadden \& Lithwick (2017),
which is a much more strictly controlled set of planetary TTVs, we find a much better correspondence with the 
predictions from our population models (the right panel in Fig. \ref{scatter}). Second, as pointed out in
Yahalomi et al. (2025), the dominant TTV term from the dynamical interaction of planets should follow a ``circus tent'' 
profile near resonances when $P^*_{\rm TTV}$ is plotted against $P_{\rm pert}/P_{\rm trans}$, where $P_{\rm pert}$
is the orbital period of perturbing planets. We reproduce 
this trend from the ${\cal M}_{219}$ model in Fig. \ref{circus}. Yahalomi's TTVs with $P^*_{\rm TTV}>10$ and 
planetary TTVs from Hadden \& Lithwick (2017) tend to follow the same profile (Yahalomi et al. 2025 and 
Fig. \ref{circus}), but the ones with $P^*_{\rm TTV}<10$ do not. Hence, the short-period TTVs are 
probably not (in general) produced by interacting planets.

\begin{figure}
\epsscale{0.8}
\plotone{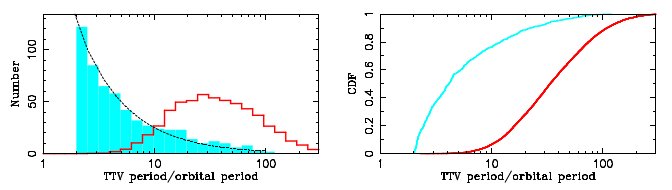}
%\epsscale{0.9}
%\plotone{fig3.eps}
\caption{The distribution of TTV periods, normalized by the orbital periods, 
for Kepler planets with significant TTVs (blue histogram and blue line; $A_{\rm TTV}>10$~min; Yahalomi et al. 2025).  
The two plots show the differential (left) and cumulative distribution functions (right).
The red lines are the distributions obtained from ${\cal M}_{219}$. {The model and measured distributions
are clearly different from each other (Kolmogorov-Smirnov and Anderson-Darling test $p$-values $<10^{-6}$).}  
The dashed line in the left panel is ${\rm Pr}(P^*_{\rm TTV}) \propto (P^*_{\rm TTV})^2$, as expected if the 
distribution for $P^*_{\rm TTV}>2$ is aliased from frequencies above the Nyquist limit (Kipping 2021).}  
\label{nyquist}
\end{figure}

Figure \ref{nyquist} shows the distribution of $P^*_{\rm TTV}$ for planets with $A_{\rm TTV}>10$ min. 
The observed and model distributions are markedly different. The observed distribution
from Yahalomi et al. (2025) peaks at the Nyquist period, $P_{\rm Nyquist}=2 P_{\rm trans}$, which is the
shortest period that can be detected with transit observations. All high-frequency terms with 
$P_{\rm TTV}<P_{\rm Nyquist}$ are aliased to $P_{\rm TTV}>P_{\rm Nyquist}$. The fact that the observed 
distribution peaks at $P_{\rm Nyquist}$ suggests that many of the observed TTV periods, especially the 
ones just above $P_{\rm Nyquist}$, can be aliases from unresolved higher frequencies. Kipping (2021)
studied this problem and found that a fully aliased profile should follow 
${\rm Pr}(P^*_{\rm TTV}) \propto (P^*_{\rm TTV})^2$, 
which we confirm to be the case (Fig. \ref{nyquist}, see the 
dashed line in the left panel). The origin of high-frequency terms in the Yahalomi's TTV catalog is unclear
(see Section 4 for discussion). 

\begin{figure}
%\epsscale{0.5}
%plotone{ttvpro.eps}\\[3.mm]
%\plotone{ttvpro2.eps}
  \epsscale{0.6}
  \plotone{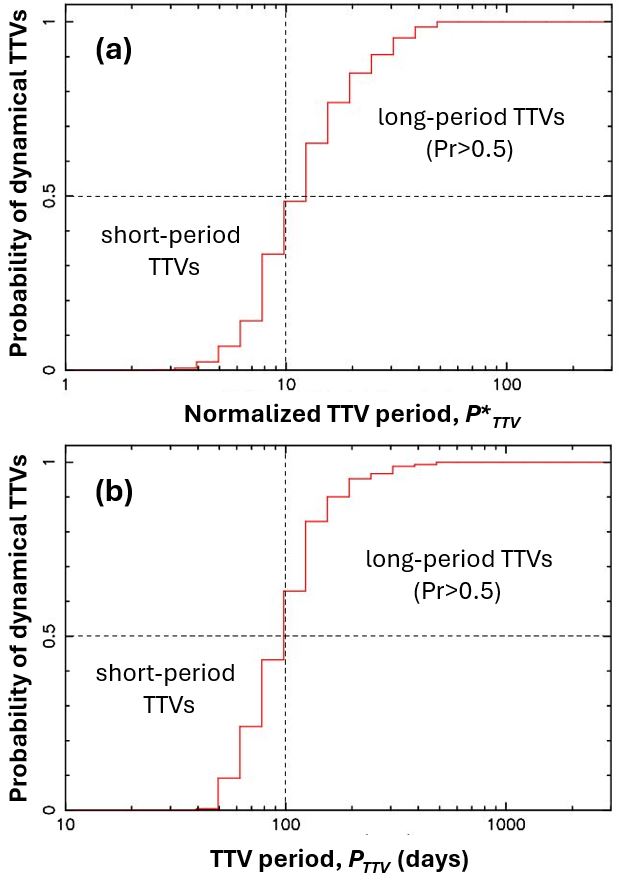}
  \caption{The probability that a TTV planet in Yahalomi et al. (2025) shows TTVs due to the dynamical interaction 
of planets. Our two component mixing model is described in Section 3.1. Here we only consider $A_{\rm TTV}>10$~min. 
Panel (a) shows the probability as a function of the normalized TTV period, $P^*_{\rm orb}=P_{\rm TTV}/P_{\rm trans}$, 
and panel (b) shows the probability as a function of $P_{\rm TTV}$. The long-period TTVs with $P^*_{\rm orb}>10$
(or $P_{\rm TTV}>100$~d) are more likely to be dynamical, whereas the short-period TTVs with $P^*_{\rm orb}<10$ 
(or $P_{\rm TTV}<100$~d) are more likely to have a different origin.}  
\label{ttvpro}
\end{figure}

{The threshold between the short-period and long-period TTVs is not strict in that not all long-period TTVs 
with $P^*_{\rm TTV}>10$ are expected to be dynamical, and some short-period TTVs with $P^*_{\rm TTV}<10$ are dynamical. 
To highlight this issue we assume that TTVs reported in Yahalomi et al. (2025) have dynamical (from planet-planet
interactions) and aliased contributions. ExoMOD is used to predict the dynamical contribution. The fraction
of model planets with dynamical TTVs as a function of normalized TTV period $P^*_{\rm TTV}$ is denoted by 
$f_{\rm dyn}(P^*_{\rm TTV})$. The aliased contribution is taken from the fit to the aliased profile 
in Fig. \ref{nyquist}. The fraction of planets with aliased TTVs, again as a function of $P^*_{\rm TTV}$, 
is denoted by $f_{\rm alias}(P^*_{\rm TTV})$. We then combine these fractional TTV contributions to compute 
the model probability that the observed TTVs are dynamical, $f_{\rm dyn}/(f_{\rm dyn}+f_{\rm alias})$,
as a function of $P^*_{\rm TTV}$ (Fig. \ref{ttvpro}). This analysis provides justification for the threshold 
between the short-period and long-period TTVs: $P^*_{\rm TTV} \simeq 10$ is where the dynamical (from
planet-planet interactions) and aliased TTVs have roughly the same contribution. For $P^*_{\rm TTV} >10$, most
TTV cases reported Yahalomi et al. (2025) should be produced by dynamical interactions of planets.} 

\subsection{Analysis of long-period TTVs}

\begin{figure}
\epsscale{0.8}
\plotone{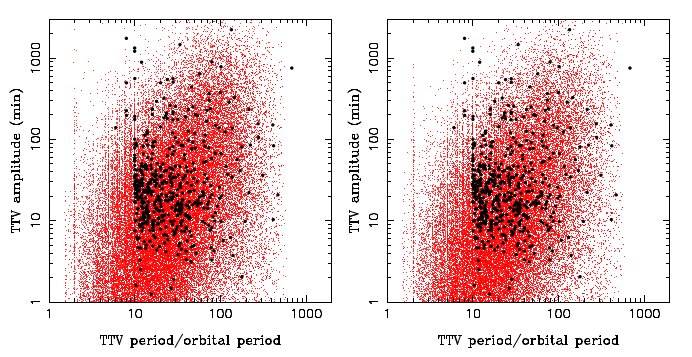}
%\epsscale{0.45}
%\plotone{fig5a.eps}\hspace*{2.mm}
%\plotone{fig5b.eps}
  \caption{The TTV amplitude and TTV period normalized by the orbital period of the transiting planet,
$P^*_{\rm TTV}=P_{\rm TTV}/P_{\rm trans}$, where $P_{\rm trans}$ is the orbital period of a transiting planet with 
the TTV period $P_{\rm TTV}$, TTVs obtained from ${\cal M}_{219}$ (left panel, Poisson) and ${\cal M}_{221}$ 
(right panel, Zipfian) are shown as red dots. The Poisson and Zipfian multiplicity models imply  
similar TTV distributions. The black dots show our new analysis of Holczer et al. (2016) TTV data that ignored 
terms with $P_{\rm TTV}/P_{\rm orb}<10$.}
\label{scatter2}
\end{figure}

We face a difficult problem of how to use Kepler TTVs if the short-period signals may not reliably linked 
to planetary interactions. We tried several options. For example, we repeated the analysis in Yahalomi et al. (2025) 
but ignored all LS harmonics with $P^*_{\rm TTV}<10$. This, of course, removed all short-period signals, but also produced 
a new pile-up just above $P^*_{\rm TTV}=10$ (Fig.~\ref{scatter2}). There is clearly an unidentified source 
of apparent short-period TTVs in the Kepler data (as discussed in the previous section), and the LS periodogram often 
opts for the shortest TTV period that is permitted to consider (Fig. \ref{ttv219}). Either we somehow model 
the TTV ramp at short periods or find better ways to filter it out from the TTV data. In this work, as a 
preliminary study of the problem, we simply adopt our new analysis and ignore $P^*_{\rm TTV}<10$. 

\begin{figure}
\epsscale{0.9}
\plotone{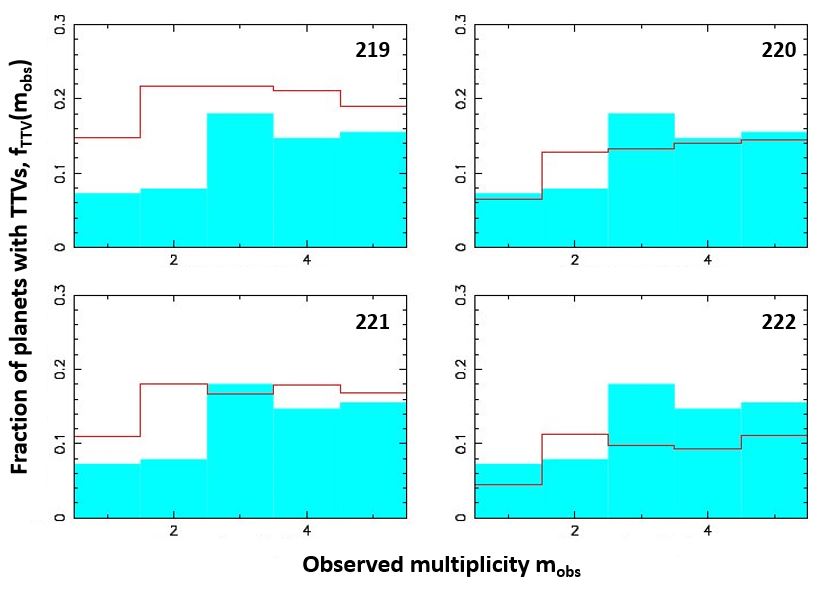}
%\epsscale{0.45}
%\plotone{model219_ttvmult.eps}\hspace*{4.mm}
%\plotone{model220_ttvmult.eps}\hspace*{4.mm}\\
%\plotone{model221_ttvmult.eps}\hspace*{4.mm}
%\plotone{model222_ttvmult.eps}
\caption{The fraction of detected planets showing significant TTVs as a function of observed 
multiplicity, $f_{\rm TTV}(m_{\rm obs})$. 
The four panels show the results for ${\cal M}_{219}$ (top left), ${\cal M}_{220}$ (top right), 
${\cal M}_{221}$ (bottom left), and ${\cal M}_{222}$ (bottom right); the red lines are the model 
TTV fractions. The Kepler TTV data from our new analysis of Holczer et al. (2016) are shown as 
the blue histograms.}
\label{ttvmult}
\end{figure}

Figure \ref{ttvmult} shows how the fraction of planets with significant TTVs -- $f_{\rm TTV}(m_{\rm obs})$ defined 
for $P^*_{\rm TTV}>10$, $A_{\rm TTV}>10$ min, and also $P_{\rm TTV} < 1000$ d to avoid complications arising from the 
Kepler baseline\footnote{If only a few transits are available for a very-long-period planet with TTVs, the LS 
periodogram cannot accurately determine the TTV period and amplitude, and typically proposes $P_{\rm TTV} 
\gtrsim 1000$ d. We do not include these cases here.}   
-- depends on the observed system multiplicity $m_{\rm obs}$. All our base models show a step with
the fraction increasing by a factor of $\sim 2$ from $m_{\rm obs}=1$ to $m_{\rm obs}=2$, $f_{\rm TTV}(2) \sim 2 f_{\rm TTV}(1)$. 
This is not reflected in the Kepler data, where we instead have $f_{\rm TTV}(2) \sim f_{\rm TTV}(1)$. Perhaps the 
the simplest solution of this problem would be if $\sim 50$\% of Kepler TTV singles were not related
to dynamical interaction between planets (Section 2.1). Still, even with this reduction, having a relatively 
large fraction of single-detected planet with long-period significant TTVs indicates that these planets 
cannot be singles. The large TTVs fraction for $m_{\rm obs}=1$ could only be explained if they have non-transiting 
companions that dynamically interact with them.

The fraction of Kepler planets with significant long-period TTVs increases by at least a factor of $\sim 2$ 
from $m_{\rm obs}=2$ to $m_{\rm obs}=3$, and shows a plateau with $f_{\rm TTV} \simeq 15$-18\% for $m_{\rm obs}=3$, 
4 and 5 (Fig.~\ref{ttvmult}). This behavior is not obtained in our population models, which instead exhibit 
a roughly constant $f_{\rm TTV}$ fraction for $2 \leq m_{\rm obs} \leq 5$. 
In Section 3.3, we consider the possibility that this problem is related to the gap complexity issue discussed 
in Paper I. Indeed, when we plot the orbital period ratio distribution of neighbor 
planets {\it only for planets showing significant TTVs}, we find that the Kepler data peaks near $P_j/P_{j-1}=1.5$, 
as expected for the dynamical interaction of planets near the 3:2 resonance (the bottom plots in Fig. \ref{ttv219}). 
This behavior is qualitatively similar to that discussed for the low gap complexity systems in Paper I 
(e.g., the KOI-82 system). 

Our Poisson multiplicity models (${\cal M}_{219}$ and ${\cal M}_{220}$) show slightly larger $f_{\rm TTV}$ fractions 
than the Zipfian multiplicity models (${\cal M}_{221}$ and ${\cal M}_{222}$). This is expected because the Poisson models with 
the average expectation $\lambda_{\rm pl}=3$-3.5 of planets (Paper I) place a great majority of planets in the multi-planet
systems, whereas the Zipfian models with the skewness $\beta_{\rm zipf}<0$ have a significant component of single planets
(that do not exhibit TTVs due to planet interactions). Overall, the Poisson models suggest somewhat larger $f_{\rm TTV}$
fractions than observed, especially for $m_{\rm obs}=2$, but this could be understood if complicating factors in the Kepler data (e.g., stellar 
activity, photometric accuracy) produced a modest loss of TTV signals in the investigated domain
($P^*_{\rm TTV}>10$, $A_{\rm TTV}>10$ min). 
    
\subsection{The 3:2 component}

\begin{figure}
\epsscale{0.8}
\plotone{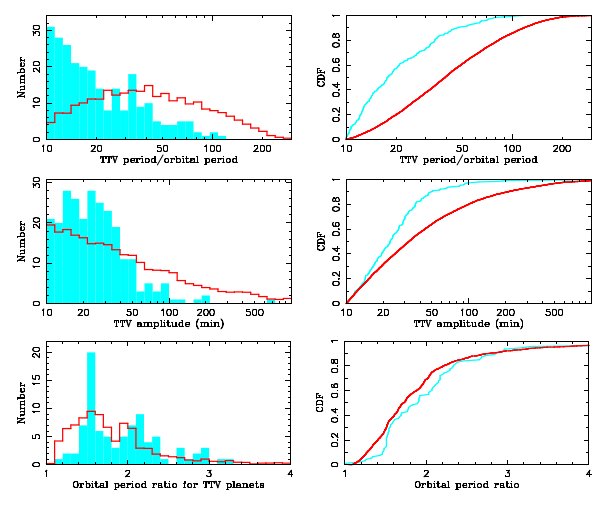}
%\epsscale{0.8}
%\plotone{fig6a.eps}\vspace*{2.mm}
%\plotone{fig6b.eps}\vspace*{2.mm}
%\plotone{fig6c.eps}
\caption{The TTV period normalized by the orbital period ($P^*_{\rm TTV}$, top), TTV amplitude ($A_{\rm TTV}$, 
middle), and orbital period ratio between the Kepler planets showing significant long-period TTVs (blue histograms 
and blue lines; $P^*_{\rm TTV}>10$, $A_{\rm TTV}>10$ min; Yahalomi et al. 2025). 
The two columns of plots
show the differential (left) and cumulative distribution functions (right). The red lines are the 
distributions obtained from ${\cal M}_{219}$.}
\label{ttv219}
\end{figure}

Figure \ref{ttv219} shows the distribution of orbital period ratios between neighbor planets 
with significant TTVs ($A_{\rm TTV}>10$ min, bottom panels). The Kepler data exhibit a relatively large fraction 
of TTV systems with $P_j/P_{j-1} \simeq 1.5$ corresponding to the 3:2 orbital resonance. ${\cal M}_{219}$ did not 
account for this feature, because there were no means built in the base models to generate orbits near resonances. 
Instead, the base models show a relatively large fraction of significant TTVs for $P_j/P_{j-1} < 1.5$ 
(Fig. \ref{ttv219}, bottom panels), as 
expected because the gravitational interaction between two planets on tightly spaced orbits is stronger,
and should produce measurable TTVs.

To address this issue, we have incorporated the 3:2 component in the auxiliary models discussed in Paper I. 
This was done by assuming that a small fraction, $f_{\rm 3:2}$, of highly correlated systems with 
$m_{\rm int}>m^*$ have orbits near the 3:2 resonance (modeled as a Gaussian distribution of period ratios
with the mean $P_j/P_{j-1} = 1.55$ and the variance 0.02). By trials and errors, we found that 
$f_{\rm 3:2} \simeq 0.1$ is best. The auxiliary fits discussed in Paper I, including ${\cal M}_{228}$,
already included the 3:2 component with $f_{\rm 3:2} = 0.1$.    

\begin{figure}
\epsscale{0.8}
\plotone{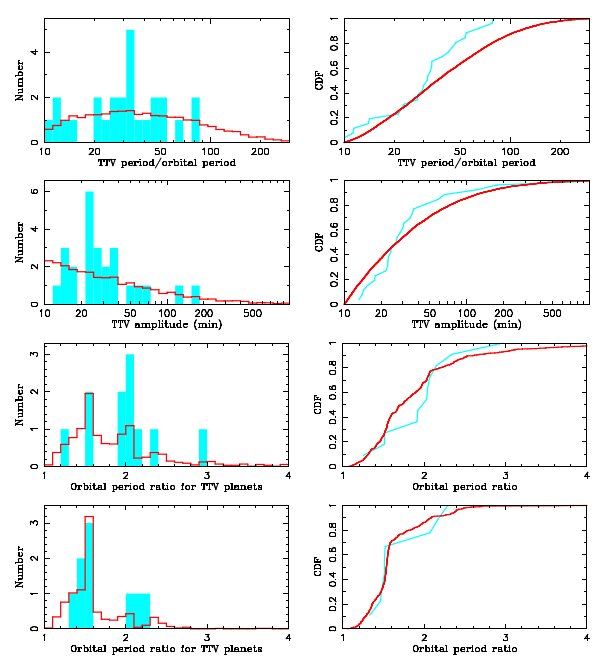}
%\epsscale{0.8}
%\plotone{fig8a.eps}\vspace*{2.mm}
%\plotone{fig8b.eps}\vspace*{2.mm}
%\plotone{lithwick3.eps}\vspace*{2.mm}
%\plotone{fig8c.eps}\vspace*{2.mm}
%\plotone{fig8d.eps}
\caption{TTVs from ${\cal M}_{228}$ ($m^*=4.5$, $f_{\rm AMD}=0.623$ and $f_{3:2}=0.1$)
are compared with Hadden \& Lithwick (2017): the TTV period normalized by 
the orbital period ($P^*_{\rm TTV}$, top row), TTV amplitude ($A_{\rm TTV}$, second row), 
and orbital period ratio between the neighbor planets 
showing significant TTVs -- the bottom two rows for $m_{\rm obs}=3$ (11 cases) and $m_{\rm obs}\geq 4$ (9 cases). 
We do not plot the results for $m_{\rm obs}=2$, where Hadden \& Lithwick (2017) only had two cases.
The two columns of plots
show the differential (left) and cumulative distribution functions (right).}
\label{lithwick}
\end{figure}

We first compare the TTV results from ${\cal M}_{228}$ with Hadden \& Lithwick (2017). This comparison 
has the advantage that the TTV data from Hadden \& Lithwick (2017) arise from a more strictly controlled 
dataset than Yahalomi et al. (2025). The disadvantage is that the statistics is small, with only the 
total of 22 robust planet pairs included here. Figure \ref{lithwick} shows that the distributions of 
normalized TTV periods and TTV amplitudes are reasonably well reproduced in ${\cal M}_{228}$
(the Kolmogorov-Smirnov or KS test probabilities are 0.15 and 0.34 for the periods and amplitudes, respectively;
the Anderson-Darling or AD test $p$-values are also $>0.1$), even though this is not particularly significant
because of the small sample size.

The comparison of orbital period ratios for planets with significant TTVs looks reasonable (the two bottom 
rows in Fig. \ref{lithwick}). Specifically, the 3:2 component included in ${\cal M}_{228}$ very closely matches 
the distribution for $m_{\rm obs} \geq 4$, where planet pairs with $P_j/P_{j-1} \simeq 1.5$ approximately 
represent 50\% of systems with significant TTVs. The 3:2 component is less prominent for 
$m_{\rm obs} \leq 3$ in ${\cal M}_{228}$, because these low multiplicity systems often derive from the systems 
with $m_{\rm int} < m^*$ and have uncorrelated orbital periods in ${\cal M}_{228}$. For $m_{\rm obs} \leq 3$, instead, 
the 2:1 component with $P_j/P_{j-1} \simeq 2$ approximately represents 50\% of TTV pairs from Hadden \& Lithwick (2017).
These differences are intriguing. They may influenced by the Kepler baseline, as it is more difficult to squeeze 
long 2:1 chains to $P_{\rm orb}<300$ d, or represent interesting clues to planet formation. For example, Bitsch 
\& Izidoro (2024) found that slower planet migration in low viscosity disks often leads to capture in the 
2:1 and 3:2 resonances, whereas high viscosity disks lead to more compact planetary configurations that 
often become unstable after the gas disk dispersal. 
 
\begin{figure}
\epsscale{0.8}
\plotone{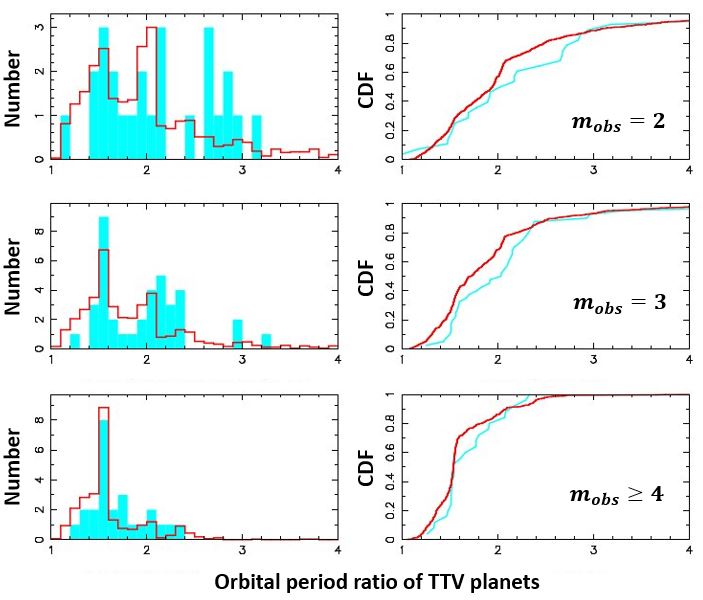}
%\epsscale{0.8}
%\plotone{afinity1.eps}\vspace*{2.mm}
%\plotone{afinity2.eps}\vspace*{2.mm}
%\plotone{afinity3.eps}
\caption{The period ratio between neighbor planets showing significant TTVs ($A_{\rm TTV}>10$ min). 
The two columns of plots show the differential (left) and cumulative distribution functions (right). 
The three rows of plots show $m_{\rm obs}=2$ (top; 28 cases in Holczer et al. 2016; blue histograms and blue lines), 
$m_{\rm obs}=3$ (middle; 11 cases) and $m_{\rm obs}\geq 4$ (bottom; 25 cases). The red lines show the results 
from ${\cal M}_{228}$ ($m^*=4.5$, $f_{\rm AMD}=0.623$ and $f_{3:2}=0.1$).}
\label{afinity}
\end{figure}

Figure \ref{afinity} shows the same analysis for Yahalomi et al. (2016). Much like with Hadden \& 
Lithwick (2017), as we progress from lower to higher
multiplicities, the 3:2 component becomes more and more prominent.
The KS test probability is 25\% for $m_{\rm obs} \geq 4$ (the bottom plot in Fig. \ref{afinity}; the AD test
$p$-value 0.3). These results, together with Fig. \ref{lithwick}, are encouraging as they give some support 
to our model for the 3:2 component in the high multiplicity systems. A similar approach could be used 
to model the 2:1 component but we leave that for future work. The statistics will need to be improved for further
progress. 

\begin{figure}
\epsscale{0.8}
\plotone{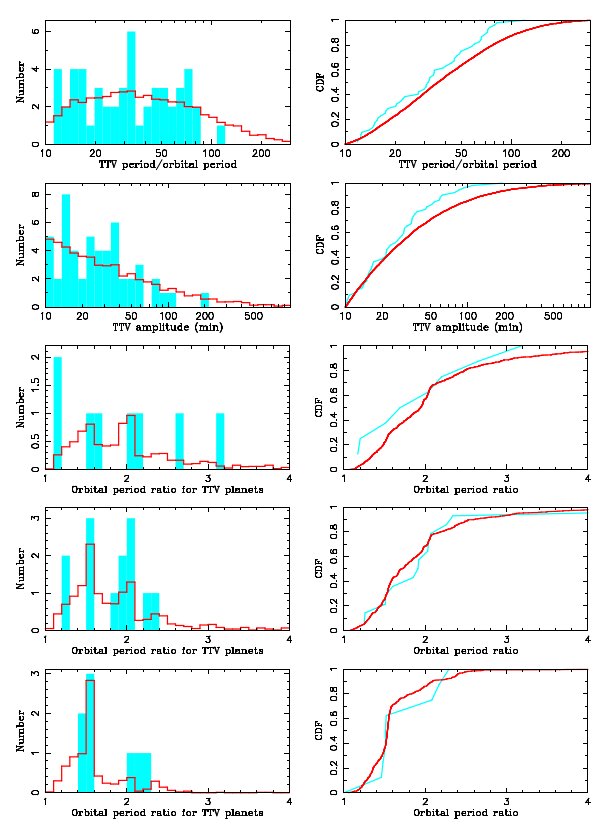}
%\epsscale{0.8}
%\plotone{fig11a.eps}\vspace*{2.mm}
%\plotone{fig11b.eps}\vspace*{2.mm}
%\plotone{fig11c.eps}\vspace*{2.mm}
%\plotone{fig11d.eps}\vspace*{2.mm}
%\plotone{fig11e.eps}
\caption{TTVs from ${\cal M}_{228}$ (Poisson multiplicity, $m^*=4.5$, $f_{\rm AMD}=0.623$ and $f_{3:2}=0.1$)
are compared with Ofir et al. (2018): the TTV period normalized by the orbital period ($P^*_{\rm TTV}$, top row), 
TTV amplitude ($A_{\rm TTV}$, second row), and orbital period ratio between neighbor planets showing significant 
TTVs -- the bottom three rows for $m_{\rm obs}=2$ (8 cases), $m_{\rm obs}=3$ (14 cases) and $m_{\rm obs}\geq 4$ 
(8 cases). The two columns of plots show the differential (left) and cumulative distribution functions (right).}
\label{ofir2}
\end{figure}

Figure \ref{ofir1} compares TTVs from ${\cal M}_{228}$ to Ofir et al. (2018). Although there is still a modest 
excess of short-periodic TTVs compared to our model predictions, the overall agreement is much better than 
for Yahalomi et al. (2025) (cf. Fig. \ref{scatter}). Figure \ref{ofir2} contrasts the TTV 
distributions in more detail. Compared to Fig. \ref{ttv219}, which indicated a large disagreement between 
${\cal M}_{219}$ and Yahalomi et al. (2025), here things look better. The normalized period distributions 
from ${\cal M}_{228}$ and Ofir et al. (2016) match each other relatively well (KS test probability 12\%). 
Since the models ${\cal M}_{219}$ and ${\cal M}_{228}$ are not much different, the improvement is mainly an 
expression of the TTV characteristics from Ofir et al. (2018). The TTV amplitudes and the orbital period 
ratios of planets with significant TTVs look good as well (KS test probabilities $p>0.1$). This indicates 
that the catalog of Ofir et al. (2018) is less affected by the aliasing problem discussed in Section 3.2. 

Related to the discussion of Fig. \ref{ttvmult} in Section 3.2, we find that including the 3:2 component in
${\cal M}_{228}$ does not help to resolve the issue with the TTV fraction, $f_{\rm TTV}(m_{\rm obs})$, which
increases by at least a factor of $\sim 2$ from $m_{\rm obs}=2$ to $m_{\rm obs}=3$. ${\cal M}_{228}$, consistently
with the base models shown in Fig. ~\ref{ttvmult}, instead predicts a roughly constant $f_{\rm TTV}$ fraction
for $2 \leq m_{\rm obs} \leq 5$. The TTV data from Hadden \& Lithwick (2017) and Ofir et al. (2018) could be
more aligned with our model predictions, but the statistics would have to be improved to meaningfully pin
down the true shape of the $f_{\rm TTV}(m_{\rm obs})$ distribution.

\section{Discussion}

\begin{figure}
\epsscale{0.8}
\plotone{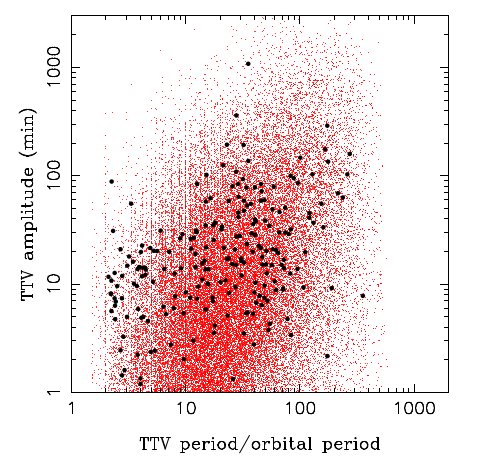}
%\epsscale{0.6}
%\plotone{fig10.eps}
\caption{The TTV amplitude and TTV period normalized by the orbital period of the transiting 
planet, $P^*_{\rm TTV}=P_{\rm TTV}/P_{\rm trans}$. The black dots is the analysis of Ofir et al. (2018). The model 
TTVs obtained from ${\cal M}_{228}$ are shown as red dots. No cut was applied here to eliminate 
the short-period TTVs.}
\label{ofir1}
\end{figure}

It would be useful to identify the source of the short-period TTV pile-up ($P_{\rm TTV}/P_{\rm orb}<10$).
Several clues have already been discussed in Section 3.1. The pile-up is unlikely to be produced by 
planet-planet interactions because: the (1) TTVs obtained from ExoMOD do not show any short-period pile-up, 
and (2) observed TTVs do not follow the circus-tent profile for $P_{\rm TTV}/P_{\rm orb}<10$ (Fig. \ref{circus}; 
Yahalomi et al. 2025). The short-period TTV pile-up also does not appear  
in the more strictly controlled TTV dataset from Hadden \& Lithwick (2017) (the right panel in Fig. 
\ref{scatter}), and is much reduced in the dataset of Ofir et al. (2018) (Fig. \ref{ofir1}).    

We found that that the pile-up may be related to aliased high frequencies ($P_{\rm TTV}/P_{\rm orb}<2$), because 
the fall off of the signal for $P_{\rm TTV}>2 P_{\rm orb}$ approximately follows the expected profile (Fig. 
\ref{nyquist}; Kipping 2021). If that's the case, the aliased signal should inherit many of the properties 
of the parent (under-sampled) signal. For example, the aliased TTV amplitude should
be equal to the unaliased TTV amplitude. We therefore infer that we are searching for an effect capable 
of producing $\sim 3$-100 min variations of transit midtimes with a short, $<2\ P_{\rm orb}$ periodicity. 

Several possibilities exist. The short-period TTVs could have some astrophysical cause. For example, they 
can be driven by the stellar activity. If so, the TTV amplitude should scale with spot activity level of the 
star, which in turn should scale with the rotational modulation amplitude. As spots can never induce TTVs 
larger than the transit duration, in this case, the TTV amplitudes would scale with transit duration. TTVs 
should also display the out-of-transit slope correlation predicted by Mazeh et al. (2015). Holczer et al. 
(2015), however, applied this method to Kepler KOI and identified only nine systems where the photometric spot 
modulation was large enough and the transit timing accurate enough to detect the correlation. Furthermore,
as rotational modulations are rarely coherent over long times, we would expect a complex Fourier spectrum 
to result.

\begin{figure}
\epsscale{0.8}
\plotone{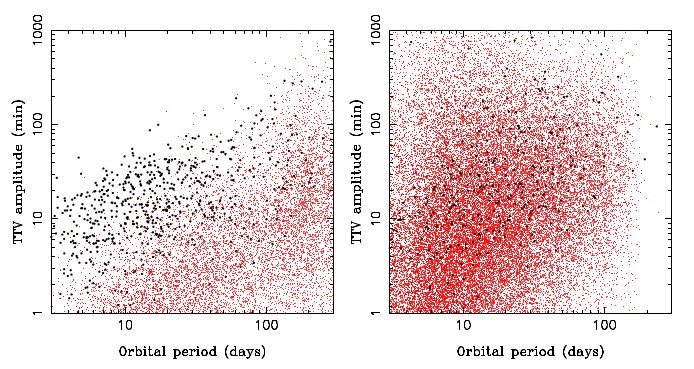}
%\epsscale{0.45}
%\plotone{fig12a.eps}\hspace*{2.mm}
%\plotone{fig12b.eps}
\caption{The orbital period and TTV amplitude of planets with $P^*_{\rm TTV}<10$ (left panel) and
$P^*_{\rm TTV}>10$ (right panel). TTVs obtained from ${\cal M}_{228}$ are shown as red dots.
The black dots are the original analysis of Holczer et al. (2016) TTV data from Yahalomi et al. (2025).}
\label{disc1}
\end{figure}

We tested several of these effects. For example, when the transit duration ($T_{\rm dur}$) is plotted against 
TTV amplitudes for the problematic cases with $P_{\rm TTV}/P_{\rm orb}<10$, we indeed find a direct correlation   
between $T_{\rm dur}$ and $A_{\rm TTV}$. As $T_{\rm dur} \propto P_{\rm orb}$, this is reflected in the correlation 
between $A_{\rm TTV}$ and $P_{\rm orb}$ as well (Fig. \ref{disc1}). Moreover, the TTV amplitudes of planets in 
the short-period pile-up are larger than the maximum TTVs expected from planetary interactions in ExoMOD.
For $P_{\rm TTV}/P_{\rm orb}>10$, instead, the $A_{\rm TTV}$-$P_{\rm orb}$ distribution of observed planets with $P_{\rm TTV}/P_{\rm orb}>10$
is fully consistent with ExoMOD (the right panel in Fig. \ref{disc1}). We did not find any significant 
correlation of $A_{\rm TTV}$ with other parameters such as the planet radius, transit depth, etc. 

\begin{figure}
\epsscale{0.8}
\plotone{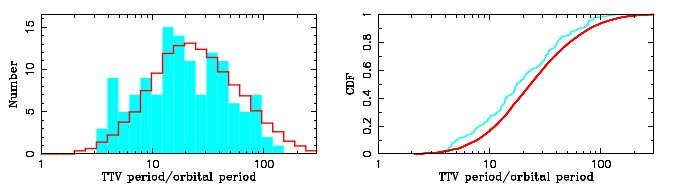}
%\epsscale{0.9}
%\plotone{fig13.eps}
\caption{The distribution of TTV periods, normalized by the orbital period of transiting planets, 
for Kepler planets (blue histogram and blue line; Yahalomi et al. 2025). Following 
the method described in Section 4, we aggressively removed all planets with TTVs that can potentially 
result from stellar rotation or its aliases. The two plots show the differential (left) and cumulative 
distribution functions (right). The red lines are the distributions obtained from ${\cal M}_{228}$.}  
\label{disc2}
\end{figure}

As a separate check, we cross-matched the stellar rotation catalog from Kamai \& Perets (2025) with the 
Holczer data. This resulted in 873 planets in the joint catalog. Subsequently, we filtered out all planets
for which there was a suspicion that the measured TTV period is equal to the stellar rotation period 
$P^*_{\rm spin}$ or one of its aliases, $P' = (1/P^*_{\rm spin} \pm m/P_{\rm orb})^{-1}$. The best results were 
obtained when we applied a generous tolerance of 70\% (probably too high), meaning that the TTV 
planets with $|P'-P_{\rm TTV}|/P_{\rm TTV}<0.7$ were removed. Only 291 planets survived this aggressive
filter. Figure \ref{disc2} compares the TTV period distribution of these planets with ExoMOD. In a vast
improvement over Fig. \ref{nyquist}, the two distributions are now comparable to each other. The 
short-period pile-up has completely disappeared, and the normalized TTV period now shows the bell-shaped 
distribution expected from ExoMOD. 

The downside of this procedure is that our aggressive filter kills a lot of TTV signal much of which is
probably unrelated to the stellar spin. It is also questionable whether the high tolerance applied here can be justified, 
because the uncertainties of $P^*_{\rm spin}$, $P_{\rm TTV}$ and $P_{\rm orb}$ are much smaller. Applying stricter
tolerance ($<50$\%), however, quickly shifts the distribution shown in Fig. \ref{disc2} to shorter periods, 
and the TTV pile-up re-appears. We leave a detailed investigation of the effect of stellar activity on
short-period TTVs for future work.  
   
Alternatively, the short-period TTV pile-up can be caused by some instrumental or methodology issues. For 
example, phasing between the long cadence (LC) Kepler data and midtransit times $t_{\rm mid}$ could induce 
high-frequency TTVs (Kipping \& Bakos 2018) that would be typically aliased to $2<P_{\rm TTV}/P_{\rm orb}<10$, 
possibly causing the observed pile-up. While the amplitude of this effect could potentially be consistent 
with the measured amplitudes, a dedicated study of this effect would be needed to test things in detail.  
 
{In our preliminary tests, we modeled the light intensity during transits by a perfectly trapezoidal profile 
(Carter et al. 2008) and ignored any photometric noise or deviations from the trapezoidal profile due to 
stellar spots or other astrophysical effects. Transits were distributed following a linear ephemeris (i.e., 
fixed orbital period, no planetary TTVs). We then convolved the photometric intensity with the 30 min 
exposure of Kepler LC observations, determined the new midtransit times $t'_{\rm mid}$, and applied
the LS periodogram to find the expected TTV amplitude and period from the phasing effect. Under the 
idealized conditions described above, this generated $A_{\rm TTV} \simeq 7$-20 min (the amplitude depends on 
the exact phasing between the transit and exposure midtimes). This could explain only a small fraction of short-period 
TTVs shown in Fig. \ref{scatter} where most short-period TTVs have $A_{\rm TTV} \gtrsim 20$ min. Also, it 
does not explain the preference of short-period TTVs for short orbital periods (Fig. \ref{disc1}). It therefore 
appears that the phasing effect cannot be fully responsible for the short-period TTVs found in Yahalomi 
et al. (2025).  The phasing effect could, however, explain most short-period TTVs in Fig. \ref{ofir1}
(Ofir et al. 2018). The aliased TTV periods from the phasing effect follow ${\rm Pr}(P^*_{\rm TTV}) \propto (P^*_{\rm TTV})^2$, 
as expected for any aliased signal from Kipping (2021), and $P^*_{\rm TTV}<10$ in a great majority of cases.}

A major limitation of this work is our inability to determine whether a model TTV signal of a
given amplitude would be detectable in the Kepler data. As a reference point, we assumed that
TTV signals with $A_{\rm TTV}>10$ min would be detected; however, it remains unclear whether this
threshold is appropriate. Establishing a robust TTV detection limit would require the development
of a realistic model incorporating the instrumental and astrophysical effects discussed above.
These effects would need to be injected into the transit midtimes generated with ExoMOD,
followed by re-computation of the TTVs to identify the parameter space in which planetary TTV
signals dominate over other TTV sources. The resulting detection limits would likely depend
on several factors, including the level of stellar activity, the methodology used to determine
$t_{\rm mid}$, and the frequencies induced by various effects. Identifying the origin of the
short-period TTV pile-up is arguably the first necessary step toward constructing a more complete model.

\section{Conclusions}

The main results of this work are summarized as follows.

\begin{enumerate}
\item The population model developed in Paper I, ExoMOD, is capable of reproducing many characteristics of 
close-in Kepler planets (Nesvorn\'y et al. 2026). This indicates certain homogeneity of the underlying population 
of Kepler planets, and probably reflects common physical process involved in their formation (e.g., disk-driven 
migration and orbital resonances). Here we used ExoMOD to statistically predict TTVs in Kepler's multi-planet 
systems. 
\item The characterization of Kepler systems could be improved from TTVs, but this is difficult to accomplish 
at this time given the unknown source of the pile-up at short TTV periods ($P_{\rm TTV}/P_{\rm orb}<10$), which
probably reflects into the distribution for $P_{\rm TTV}/P_{\rm orb}>10$ as well. We found that the short-period 
TTV pile-up in Yahalomi et al. (2025) is most likely a consequence of aliased high frequencies 
($P_{\rm TTV}/P_{\rm orb}<2$), potentially related to some unresolved issue with the methodology. Other possible 
causes include crossing of rotating stellar spots, exomoons (e.g., Kipping 2021), active stars with high 
frequency variability, etc. 
\item The frequent occurrence of significant long-period TTVs ($P_{\rm TTV}/P_{\rm orb}>10$) among many Kepler 
singles suggest that these planets have non-transiting companions. ExoMOD was used to estimate that the fraction 
of systems with significant long-period TTVs ($P_{\rm TTV}/P_{\rm orb}>10$ should increase by a factor of $\sim 2$ 
from the observed multiplicity $m_{\rm obs}=1$ to $m_{\rm obs}=2$, but this is not reflected in the Kepler data
from Yahalomi et al. (2025). It could mean that $\sim 50$\% of long-period TTVs obtained in Yahalomi 
et al. (2016) for single-detected planets are unrelated to dynamical perturbations.         
\item We identified a surge of significant long-period TTVs ($P_{\rm TTV}/P_{\rm orb}>10$) for Kepler planets with 
$m_{\rm obs} \geq 3$ that is unexplained in our models -- our models instead indicate a constant TTV fraction
for $2 \leq m_{\rm obs} \leq 5$. The surge could be related to a class of Kepler planets with highly correlated 
orbital period ratios near the 3:2 and 2:1 resonances, but including the 3:2 component in ExoMOD did not help
to explain it. 
\item The 3:2 component, which was included in our auxiliary models ($f_{\rm 3:2}=0.1$ for $m^*=4.5$; i.e., in 10\% 
of high-multiplicity systems with $m_{\rm int} \geq 5$) helped to explain the orbital period ratio distribution
of planets with significant TTVs (amplitudes $A_{\rm TTV}>10$ min; Figs. \ref{afinity} and \ref{ofir2}). This 
indicates that $\sim 10$\% of high multiplicity systems formed and survived near the 3:2 resonance chains.    
\end{enumerate} 

ExoMOD can be used to produce statistical samples of intrinsic and Kepler-detected planetary systems. Ten samples 
from the ${\cal M}_{228}$ model, generated with different random seeds, can be found at 
{\tt www.boulder.swri.edu/\~{}davidn/ExoMOD/}. The catalog includes the TTV periods and TTV amplitudes of 
model planets. Additional model information is available upon request. {We thank the anonymous reviewer 
for their insightful comments on the submitted manuscript.}

\acknowledgements

\begin{center}
{\bf Acknowledgments} 
\end{center} 
\vspace*{-3.mm}
The simulations were performed on the NASA Athena Supercomputer. We thank the NASA NAS computing division for 
continued support. D.N.'s work was funded by the NASA XRP program. D.A.Y's work is supported by a Juan Carlos 
Torres Postdoctoral Fellowship At the Massachusetts Institute of Technology.

\section{Appendix A: Comparison with Zhu et al. (2018)}

{Zhu et al. (2018) used an approximate TTV amplitude indicator based on orbital period 
ratios of model planets. Specifically, they selected two planets in the same (model) system that are the 
closest to each other, found a nearby resonant term, and computed the corresponding super-period 
for the selected pair. Their TTV algorithm subsequently assumed that the TTV period is equal to the selected 
period, which may not be the case for many systems. For example, this procedure does not factor in whether 
the two planets are massive enough to produce measurable TTVs. In addition, the dominant TTVs may not be 
produced by the closest planets; they may instead be generated by more massive perturbers on more distant 
orbits. For comparison, in ExoMOD, we accurately calculate the period and amplitude of the dominant TTV 
term with numerical simulations. 

Zhu et al. (2018) estimated whether any planet's TTVs can be detectable or not but did not discuss 
the distribution of TTV amplitudes/periods. Their underlying model for planetary systems approximated 
the multiplicity distribution following Tremaine \& Dong (2012), which is a more general approach than 
the one adopted in ExoMOD (we used the Poisson and Zipfian distributions; Paper I). Their forward modeling adopted 
the distribution of transit parameter $\epsilon = R_*/a$ from Tremaine \& Dong (2012), Rayleigh distribution 
of orbital inclinations with the scale parameter $\sigma_i \propto m_{\rm int}^\alpha$, where $m_{\rm int}$ is the
intrinsic multiplicity and $\alpha$ is a model parameter, stability criteria from Deck et al. (2013), and 
transit detectability based on the geometric condition of transit. 

For comparison, ExoMOD infers orbital eccentricities and inclinations from the AMD-based criteria (Paper I,
He et al. 2020). The transit detectability in ExoMOD employs accurate criteria from Kipping \& Sandford (2016).
ExoMOD accurately matches many different characteristics of Kepler planets. We therefore have confidence
that the underlying models are accurate representations of the Kepler data.

Zhu et al. (2018) included TTVs in the likelihood term to optimize their models. This differs from our strategy
in ExoMOD, where we instead first calibrate the model on the DR25 catalog (TTV constraints {\it not} used in 
the likelihood term), then use the calibrated models to predict the distribution of TTVs, and finally compare 
the results with the actual TTV data (Hadden \& Lithwick 2014, 2017; Holczer et al. 2016; Ofir et al. 2018; 
Yahalomi et al. 2025). We believe that this approach is reasonable because measured TTVs may have a number of
astrophysical and/or instrumental causes, which will need to be understood before TTVs can be used to 
directly calibrate population models.}

\begin{table}
\centering
{
\begin{tabular}{lcccc}
\hline \hline
Model           & Multiplicity      &  $f_{\rm AMD}$  & $m^*$    & $\Delta \ln {\cal Z}$   \\                      
\hline                    
${\cal M}_{219}$ & Poisson           &  1.0       & --        & -0.5            \\ 
${\cal M}_{220}$ & Poisson           &  0.3       & --        & -3.6            \\ 
${\cal M}_{221}$ & Zipfian           &  1.0       & --        &  0            \\ 
${\cal M}_{222}$ & Zipfian           &  0.3       & --        & -2.9            \\
\hline
${\cal M}_{230}$ & Poisson           &  --       & 2.5        & -17.4         \\ 
${\cal M}_{229}$ & Poisson           &  --        & 3.5       & -7.1          \\ 
${\cal M}_{228}$ & Poisson           &  --        & 4.5     &  0            \\ 
${\cal M}_{231}$ & Poisson           &  --        & 5.5       & -17.1          \\ 
\hline \hline
\end{tabular}
}
\caption{A summary of our models from Paper I (Nesvorn\'y et al. 2026). 
The columns are: the (1) model designation, (2) adopted intrinsic multiplicity 
distribution, (3) fraction $f_{\rm AMD}$ of the critical AMD distributed among planets, (4) 
multiplicity transition $m^*$ to highly correlated orbital radii, and (5) Bayes factor 
difference relative to the best model in each category. The four models on the top 
are our base models with six priors and two likelihood arrays.
The four models in the bottom part of the table are our auxiliary models with seven priors 
and three likelihood arrays. See Paper I for detailed information. The auxiliary model 
${\cal M}_{228}$ is strongly favored over all other auxiliary models.}
\end{table}

\clearpage

\clearpage

\clearpage

\clearpage

\clearpage

\clearpage

\clearpage

\clearpage

\clearpage

\clearpage

\clearpage

\clearpage

\clearpage

%\clearpage
%\begin{figure}
%\epsscale{0.6}
%\plotone{model228_ttvscatter2.eps}
%\caption{A summary plot for the excess of short-period TTVs in the Kepler data. The plot shows the TTV amplitude 
%and TTV period, both normalized by the orbital period of the transiting planet, $A^*_{\rm TTV}=A_{\rm TTV}/P_{\rm trans}$ 
%and $P^*_{\rm TTV}=P_{\rm TTV}/P_{\rm trans}$. The black dots is the original analysis of Holczer et al. (2016) 
%data from Yahalomi et al. (2025). The model TTVs obtained from ${\cal M}_{228}$ are shown as red dots. The upper 
%bounding (diagonal dashed) curve of model TTVs follows $A^*_{\rm TTV} \propto (P^*_{\rm TTV})^2$, as expected from 
%Lithwick et al. (2012) for significant free eccentricities of planetary orbits near orbital resonances.}
%\label{scatter3}
%\end{figure}

%\clearpage
%\begin{figure}
%\epsscale{0.8}
%\plotone{chains.eps}
%\caption{}
%\label{chains}
%\end{figure}

\end{document}